# Boiling Flow Visualization Experiment Scaled to Divertor Cooling Conditions

*Aljoša Gajšek[a], Gregor Kozmus[a], Matej Tekavčič[a],

Marianne Richou[b], Boštjan Končar[a]

[a]Jožef Stefan Institute, Reactor Engineering Division, Ljubljana, Slovenia;

[b]CEA, IRFM, F-13108 Saint-Paul-Lez-Durance, France

A dedicated experiment FEDORA (Fusion Experiment for Divertor Optimization Research Applications) has been constructed to visualize flow boiling at scaled fusion divertor conditions. Its main purpose is to obtain the information on local boiling parameters, needed for the development of mechanistic boiling models in Computational Fluid Dynamics (CFD) codes. Current boiling models, although grounded on realistic physical principles, still rely heavily on empirical correlations, particularly for boiling parameters such as nucleation site density, bubble departure diameter, and bubble departure frequency. Due to the harsh operating conditions in the water-cooled divertor plasma-facing components (high heat fluxes over 10 MW/m$^2$, mean flow velocity of about 9 m/s), visualization of boiling in realistic cooling channels is not feasible. The proposed experiment addresses this challenge by employing a surrogate coolant, characterized by lower latent heat and saturation temperature. By analyzing key dimensionless numbers, the aim is to replicate boiling conditions to be similar to those in coolant channels of upcoming divertor designs. Distinctive features of the proposed experimental setup include visually transparent boiling channel, infra-red transparent window and a thin electrically heated foil. This setup allows simultaneous high-speed and high-resolution recording of boiling phenomena in the flow and measurement of temperature distribution on the heated surface. In this work, the design, similarity analysis and the first measurement results are presented, proving the concept of the visualization approach to understand the boiling heat transfer in divertor cooling channels.

Keywords: Flow Boiling, High-Heat Flux, Divertor Cooling Conditions, Visualization Experiment, Boiling Parameters

## 1. Introduction

Flow boiling is a highly efficient heat transfer mechanism, capable of dissipating substantial amounts of energy at a relatively small rise in surface temperature. For this purpose, boiling phenomena has been extensively studied for the analyses of nuclear fission reactors, leading to the development of advanced models for Multiphase Computation Fluid Dynamics (MCFD) simulations [1], that can model boiling phenomena at local flow conditions. These models in general are not suitable to simulate boiling in fusion divertors [2], where the operating conditions are more demanding. Incident heat fluxes on plasma facing surfaces can reach over 10 MW/m$^2$ [3] during steady-state operation, an order of magnitude higher than on the surface of fuel rods of fission reactors [4], requiring higher mean flow velocity and subcooling of the coolant to prevent the onset of boiling crisis, the combination which the current models were not designed to simulate.

The baseline of mechanistic boiling models was established with a three-heat flux partitioning model [5] developed by the Rensselaer Polytechnic Institute (RPI). The RPI model was originally developed for pool boiling and then adapted for flow boiling [6]. Recent studies include more detailed modelling of boiling heat transfer mechanisms, making the model more applicable for forced convective boiling flows [7]. Importantly, all variations of the model depend on the modelling of physically observable boiling parameters, the more important being the nucleation site density, bubble departure frequency, and bubble departure diameter. The underlying physical mechanisms of boiling occur on too small time and spatial scales to be directly resolved by numerical simulations, so the boiling parameters need to be experimentally determined and empirically modelled for credible use in MCFD simulations [8].

A significant challenge in improving MCFD models for simulating actively cooled divertor targets is the lack of experimental data that closely replicates the extreme conditions within divertor cooling channels. In this study, the water-cooled divertor conditions in the Wendelstein 7X [9],[10] are addressed: mean velocity ~ 9 m/s, incident heat flux ~ 10 MW/m$^2$ and liquid subcooling ~ 150 °C). Previous visualization experiments [11], have provided valuable insights into boiling phenomena but were conducted at lower heat fluxes, flow velocities, and subcooling levels. Therefore, the empirical correlations derived from these studies are not directly applicable to divertor cooling systems.

While realistic water-cooled divertor mock-ups can be tested under extreme thermal loads in the High Heat Flux (HHF) facilities like HADES [12] or GLADIS [13], it is not possible to observe the boiling flow in the cooling channels due to the material opacity and harsh operating conditions. Only the temperature of the heat-loaded surface of the mock-up can be measured. In addition, the construction of mock-ups and their testing in HHF facilities is a costly and time-consuming process. Due to the lack of experimental data and the poor understanding of boiling under the extreme conditions of fusion divertors, MCFD simulations cannot be used to predict the wall temperature to a reasonable degree of accuracy.

---

*Corresponding author: Aljosa.Gajsek@ijs.si

This knowledge gap is addressed by proposing a new Fusion Experiment for Divertor Optimization Research Applications (FEDORA), with a transparent test section. Since high heat fluxes are not manageable in a laboratory set-up, a surrogate coolant is used. A dimensional analysis has been performed using key dimensionless numbers to scale down the conditions typical for the fusion divertor. The distinctive features of our experiment include a visually transparent boiling channel, an infrared transparent window, and electrically heated stainless-steel foil. This configuration allows simultaneous high-speed visualization of the boiling flow and measurement of the temperature distribution on the heated surface. In this study, the design, similarity analysis, and the first results of the new experimental setup are presented.

## 2. Design

The relevant thermo-hydraulic conditions for water-cooled divertor mock-ups cannot be easily replicated in the controlled laboratory environment. Heat fluxes of more than 10 MW/m$^2$ cannot be achieved by electrical resistance heating, instead electron or ion beam guns operating in vacuum chambers are required. Additional constrains arise from a limited choice of transparent materials that can withstand high temperatures, temperature gradients, and pressures. To overcome these limitations, a surrogate liquid is used in the FEDORA experiment to relax the operating conditions, particularly the high heat flux. The exact similarity with the boiling flow in a water-cooled divertor is difficult to achieve due to the complexity of the two-phase flow structures, however the key phenomena can be reasonably approximated [14] by matching the dimensionless numbers, as presented in Section 3.

The main feature of the FEDORA experiment is a visually transparent test section, allowing optical access from all directions. This feature is crucial for high-speed visualization of boiling phenomena, providing valuable insights into bubble dynamics, nucleation site distribution, and flow patterns under conditions relevant to fusion divertor cooling channels. A square test section was selected to mitigate image distortion due to differences in the light reflection index between the air, the channel wall, and the coolant. An ultra-thin conductive foil is used for electrical heating to precisely apply power through the heated surface.

### 2.1 Test section

The test section consists of a horizontal flow channel with a square internal cross-section of 10 mm x 10 mm, keeping the same hydraulic diameter ($d_h$) as the real divertor coolant channel designs (see Fig. 1). The total length of the test section is 1400mm, with 1000 mm (100$d_h$) long inlet part upstream the heating part (HP) to ensure a fully developed turbulent flow (see Fig. 2).

The heated part of the test section with main design features is shown in Fig. 1. The 15 mm thick channel walls are constructed from optically transparent polycarbonate (PC) plastics, chosen for its transparency in the visible spectrum, mechanical strength, and temperature resistance up to 140 °C.

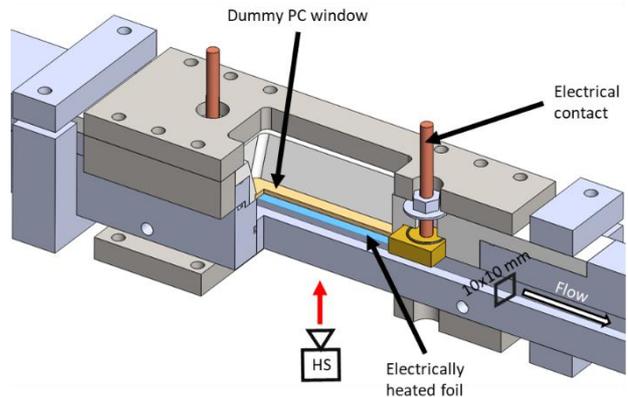

Fig. 1: Heated part (HP) of the test section, with key design features and HS camera position.

The polycarbonate is not transparent for the long-range infra-red (IR) spectrum (7.5-12 µm), in which the sensor of the High Speed (HS) thermal camera operates. For this reason, a zinc sulphide (ZnS) IR transparent window is planned to be installed in the near future. For preliminary tests, a dummy window made of polycarbonate is used in this study.

### 2.2 Heating system

The coolant is heated through a 100 mm long and 9 mm wide stainless-steel foil with a thickness of 5 µm, mimicking the one-sided heating in a divertor. The foil is in direct contact to the fluid on one side and glued to the window on the other side using a two-side black adhesive tape, ensuring high emissivity of the foil. The foil is slightly narrower than the test section to avoid corner effects. At each end, the foil is glued with an electrically conductive adhesive to the thick copper contacts connected to the battery, that ensures a constant electrical potential of 24 V on the heating foil, with a planned upgrade to 48 V. The DC current through the foil is regulated by an electric load connected in series. This configuration allows a heating power up to ~500 W (1000 W with upgrade), resulting in a heat flux up to ~500 kW/m$^2$ (1 MW/m$^2$ with upgrade) on the surface of the foil. Considering the downscaling to the experimental conditions (see Table 2 in Section 3 for more details), an experimental heat flux of 620 kW/m$^2$ corresponds to the operating heat flux of 10 MW/m$^2$ in the W7X divertor.

### 2.3 Thermal losses

To evaluate the thermal losses ($q_{loss}$), let's first conservatively estimate the wall temperature using Chen's corelation[15] to $T_{wall} = 90$ °C, or 50°C superheat at 2.5 bar, as a very high wall superheat in a nucleate boiling regime.

The heat dissipates though the PC window of thickness d = 2 mm and thermal conductivity k = 0.23 W/mK into the ambient air at $T_{air} = 20°C$. Assuming a rather high $h_{air}$= 10 W/m²K, the heat losses though the PC window and air can be estimated as

$$\frac{q_{loss}}{q} = \frac{(\frac{d}{k} + \frac{1}{h_{air}})^{-1}(T_{wall} - T_{air})}{q} \approx 0.001, \quad (1)$$

where q is the total electric heat flux, calculated as $q = UI/S = 620$ kW/m² (look at Table 1), where $U$ is voltage drop on the foil, $I$ is the electrical current, and $S$ is the area of the foil.

The heat losses due to radiation are estimated as negligible due to the low temperature difference between the ambient and the heated foil.

**2.4 Thermal hydraulic loop**

The thermo-hydraulic loop is presented in Fig. 2. A powerful pump (EDUR pbmx 201 e6.1), capable of reaching mass flows up to 1 kg/s, circulates the coolant (Honeywell Enovate R245fa) around the closed system. The flow rate through the test section is regulated by a bypass regulating valve and accurately measured with a Coriolis flow meter (Emerson Elite CMFS040M). The coolant is preheated in a heat exchanger to a stable temperature before entering the test section. Pressure and temperature are measured before and after the test section. After the test section, a condenser dissipates the heat and removes any remaining vapor bubbles from the coolant before it re-enters the pump. The pressure in the loop is regulated by an air bladder pressurizer.

Prior to operation, the loop is purged by nitrogen gas to remove the humid air from the system. Next, the nitrogen is removed by a vacuum pump, creating a medium vacuum level (100 Pa). Ultimately, the loop is filled with pure R245fa coolant, using a refrigerant recovery machine (Promax Ecomax-E).

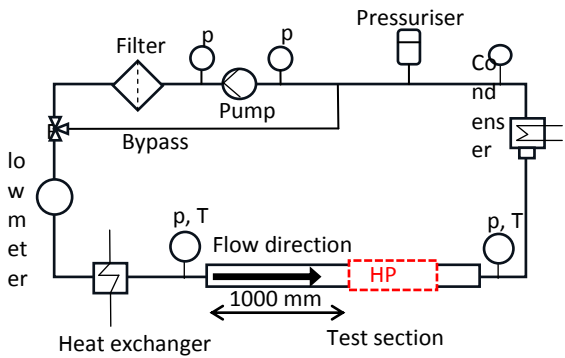

Fig. 2: Thermal hydraulic loop of the FEDORA experiment.

**2.4 Optical measurement equipment**

The experiment is designed for simultaneous measurements from three different directions using high speed cameras. For the preliminary measurements presented in this study, only a high-speed (HS) camera with a bottom view was used (see Fig. 2). The visual light high-speed camera (Phantom VEO 710L) is installed bellow the heated foil and provides a bottom-up view on the heated surface.

## 3. Similarity analysis

While similarity between different systems is ideally assessed using dimensionless groups from conservation laws, boundary conditions, and constitutive relations, this is rarely feasible due to the model's complexity and the number of unknowns involved. Instead, flow similarity analysis is a more practical tool for identifying dominant effects and constructing simplified formulations. These simplified models, though approximate, often yield valuable engineering insights [16].

### 3.1 Dimensionless numbers

The thermal-hydraulic conditions in fusion divertor are characterised by high velocity, high heat-flux and high subcooling of the flow. Our study is based on the similarity analysis approach used in the DEBORA experiments [13], which is a widely recognized methodology in boiling flow studies under scaled conditions [17], [18], [19]. The following phenomena in the flow are therefore addressed by similarity analysis: turbulence (velocity), boiling on the heated surface (heat flux), condensation (subcooling), system pressure (density ratio). The final flow feature considered in similarity analysis is the interfacial area, which depends on the bubble size and shape. The following dimensionless numbers are used to scale these phenomena:

- Liquid Reynolds number preserves the similarity in flow conditions and turbulence

$$Re = \frac{\rho_l u\, d_h}{\mu_l}, \quad (3)$$

where $\rho_l$ is liquid density, $u$ is mean flow velocity, $d_h$ is hydraulic diameter and $\mu_l$ is liquid dynamic viscosity.

- Boiling number (*Bo*) is used scale the corresponding heat flux on the heated surface

$$Bo = \frac{q}{Gh_{lv}}, \quad (4)$$

where $G$ is mass flux, and $h_{lv}$ is the latent heat of the evaporation.

- Jacobs number (*Ja*) at the inlet determines the similarity of inlet subcooling conditions and condensation potential in the bulk

$$Ja = \frac{\rho_l c_{p,l}(T_l - T_{sat})}{\rho_v h_{lv}}, \quad (5)$$

where $c_{pl}$ is the specific heat capacity at the constant pressure, $T_l$ is liquid temperature, and $\rho_v$ is the vapour density.

- Similarity in system pressure is determined by the vapor to liquid density ratio

$$\rho^* = \frac{\rho_v}{\rho_l}, \tag{6}$$

- Similarity of flow regime is captured by the Weber number (We)

$$\mathrm{We} = \frac{\rho_l u^2 d_h}{\sigma}, \tag{7}$$

where $\sigma$ is the surface tension. Usually the bubble diameter is used as a characteristic length in the We number, but as the bubble size in the flow is not known a priori under given operating conditions, the hydraulic diameter of the channel $d_h$ is used as the characteristic size instead, based on the approach used in the DEBORA experiment [14].

### 3.2 Coolant selection

Using the presented dimensionless numbers, the following function was proposed

$$\xi = \sum_{i=1}^{N}\left(1 - \frac{X_{s,i}}{X_{f,i}}\right)^2, \tag{8}$$

where the subscript $f$ stands for water under fusion conditions and $s$ for surrogate coolant under scaled-down laboratory conditions. Based on the limitations of our thermal-hydraulic loop and test section, some basic constrains were employed. The maximum pressure limit was set to $p_{\max} = 4$ bar, mean flow velocity $u_{\max} = 5$ m/s, inlet temperature $T_{\text{inlet}} = 10\ °C$, heating power $P_{\max} = 2000$ W.

The minimum function value $\min(x(p, u, T_{\text{inlet}}, P))$ over a given range of operation conditions was calculated for 122 different coolants in the CoolProp library [20]. The best results were obtained for some carbohydrates with $\xi = 0.09$ for N-hexane. N-hexane, together with other flammable, explosive, and toxic coolants was avoided due to safety concerns and therefore, the selected coolant of choice is R245fa, with $\xi = 0.36$. Comparison between the water-cooled divertor conditions in W7X [9] and the laboratory values are summarized in Table 1, and the dimensionless ratios are presented in Table 2. It should be noted that the similarity of the system pressures ($\rho^*$) and inlet subcooling ($Ja$ number) cannot be fully matched. Due to the limitations of the current cooling system, an inlet temperature below 10°C cannot be achieved, and lower pressures also have to be avoided due to the possibility of cavitation in the pump.

Table 1: Operating conditions for the W7X divertor and the FEDORA experiment with R245fa.

| Condition | W7X Divertor (water) | FEDORA (R245fa) |
|---|---|---|
| $p_{\text{abs}}$ [bar] | 10 | 2.5 |
| $u$ [m/s] | 9 | 3.6 |
| $q$ [kW/m$^2$] | 10 000 | 620 |
| $T_{\text{inlet}}$ [°C] | 30 | 10 |

Table 2: Dimensionless numbers for the W7X divertor cooling conditions and the FEDORA experiment.

|  | W7X divertor (water) | FEDORA (R245fa) | Ratio |
|---|---|---|---|
| Re | 100 000 | 100 000 | 1 |
| We | 15 000 | 15 000 | 1 |
| Bo | 4.4 e-4 | 4.4 e-4 | 1 |
| Ja | 45.7 | 14.8 | 0.3 |
| $\rho^*$ | 5.1 e-3 | 11.1 e-3 | 0.5 |

## 4. Results

### 4.1 Preliminary experiments

To test the performance of the test section design and the measuring equipment, preliminary tests were conducted using only the bottom high-speed camera, and a dummy window. The experiments were conducted at a given velocity, with varying power, and at a given power, varying the velocity. All experiments were conducted at constant inlet temperature of 10°C and an absolute pressure of 2.5 bar. The experimental matrix is presented in Table 3.Table 3: Experimental test matrix

|  | Re [ ] | Bo [ ] | $u$ [m/s] | $q$ [kW/m$^2$] |
|---|---|---|---|---|
| Fixed inlet velocity | 56 000 | 1.9 e-4 | 2 | 100 |
|  | 56 000 | 2.9 e-4 | 2 | 150 |
|  | 56 000 | 3.8 e-4 | 2 | 200 |
|  | 56 000 | 4.8 e-4 | 2 | 250 |
|  | 56 000 | 5.5 e-3 | 2 | 300 |
| Fixed heat flux | 28000 | 7.3 e-4 | 1 | 200 |
|  | 56000 | 3.8 e-4 | 2 | 200 |
|  | 84000 | 2.5 e-4 | 3 | 200 |

For each of the tests, steady-state flow conditions had to be achieved first. Afterwards, the images of the boiling flow on the heated foil were taken using the visual HS camera (frame rate 10 000 fps, exposure time 50 μs). Only the middle 5 cm segment of the heated part of the channel was observed, in order to improve the spatial resolution and enable the detection of the smallest bubbles. These recordings are presented in Figs. 4 and 5. Fig. 4 shows the images at constant inlet velocity and variable heat flux, while Fig. 5 shows the flow images at a constant heat flux of 200 kW/m$^2$ and a variable inlet velocity. As expected, it can be seen that increasing Bo by either increasing the velocity or the power, leads to increase the number of bubbles. To further study the bubble populations a more systematic approach using a bubble detection algorithm has been developed.

### 4.2 Bubble detection

The bubble detection algorithm, developed in MATLAB using the Image Processing Toolbox, identifies individual bubbles in the boiling flow. First, the raw image is preprocessed and converted to a binary format, isolating bubbles from the background based on intensity thresholds. Bright, reflective background regions, that are caused by detachment of foil sections, are then removed to ensure accurate bubble detection. In the final stage, bubbles are detected by identifying discrete white structures. For each identified structure, the algorithm calculates an equivalent diameter, defined as the diameter of a circle with an area equal to that of the detected structure. The flowchart of this algorithm is presented in Fig. 3 and the final results can be observed in Fig. 6. The raw image taken by the HS camera is shown on the top and the processed image of the bubbles is shown underneath.

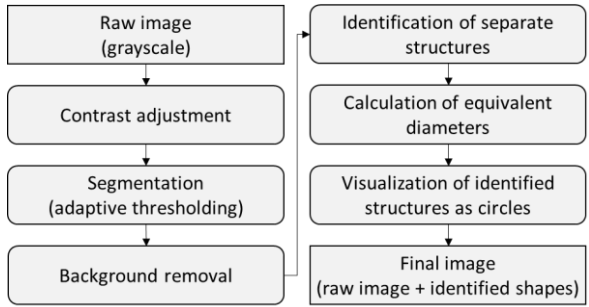

Fig. 3: Flowchart of the image processing algorithm.

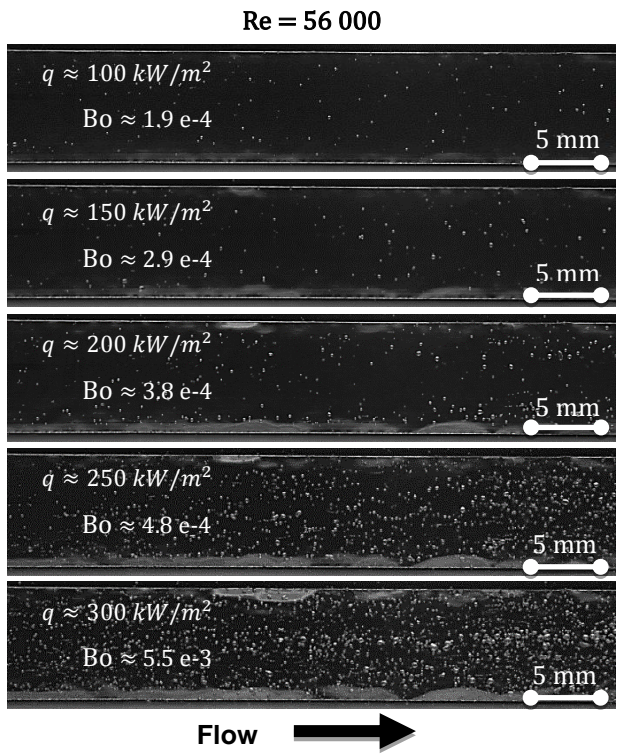

Fig. 4: A snapshot of boiling on the heated foil at a fixed Re and different heat fluxes.

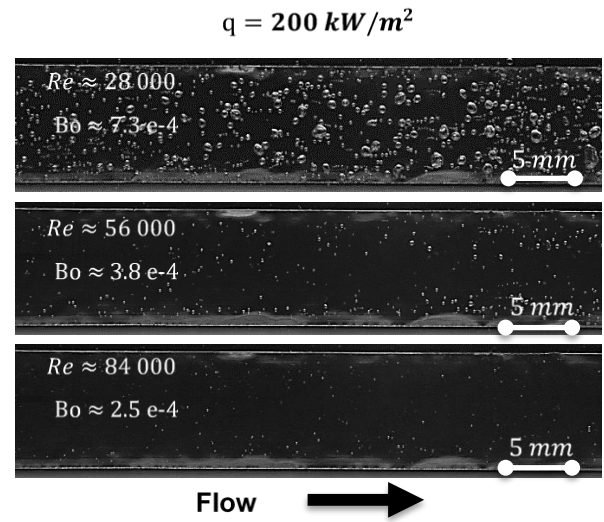

Fig. 5: A snapshot of boiling on the heated foil at a fixed heat flux and different Re numbers.

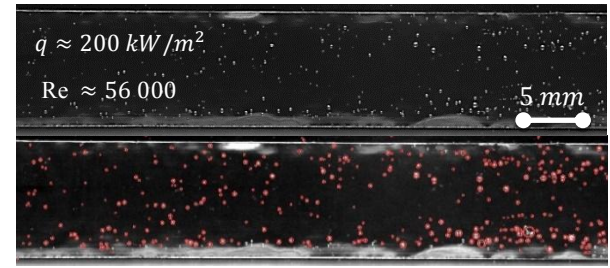

Fig. 6: An example of bubble detection algorithm (in red the bubbles detected).

**4.3 Bubble size distribution**

Using the mentioned algorithm, on the images presented in Figs. 4 and 5, the bubble size distributions were obtained, and are presented in Figs. 7 and 8. On x-axis the bin sizes (bubble diameter range) in mm are presented and on y-axis the number of bubbles per wall area of the observed segment of the channel in [$m^{-2}$] are shown.

In Fig. 7, the bubble size populations at constant heat flux (200 kW/m²) are presented for three different inlet velocities. In the first bin there is approximately the same number of bubbles for all velocities. In second bin the 1 m/s and 2 m/s are both quite similar, while there is a significant drop in number of bubbles for the 3 m/s. This trend continuous for the higher bins, with only the lowest inlet velocity of 1 m/s having a significant bubble population of bubble sizes above 0.5 mm. The shift in population could be either the result of, bigger bubbles forming on the nucleation sites, larger number bubbles in total leading to more coalescence, or a longer time that bubble spends on the heated surface leading to higher thermal growth while sliding.

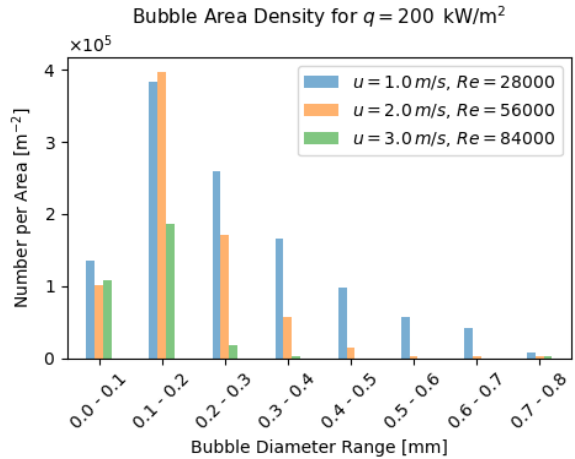

Fig. 7: Bubble size distribution for constant heat flux on the heated surface at three different inlet velocities.

In Fig. 8, the bubble size populations at constant velocity of 2 m/s are presented. For lower heat fluxes (100 kW/m², 150 kW/m², and 200 kW/m²), the number of bubbles in the first bin is approximately the same, but rises significantly at 250 kW/m², and 300 kW/m². In larger bins this trend becomes even more pronounced, as lower heat fluxes keep populations smaller and smaller. This can be explained due by a combination of more bubbles being created at higher heat fluxes, larger average size of the bubbles forming on the nucleation sites due to larger growth force acting on the nucleating bubble, and more active coalescence due to the higher bubble collision frequency.

To better understand the strength and nature of these effects, a more detailed investigation is required. This should involve tracking the evolution of bubbles from their nucleation sites along the entire heated section and assessing how bubble dynamics influence the heat transfer. Such an analysis will help to clarify the contribution of mechanisms like bubble sliding, bubble growth, and bubble size at the nucleation to the boiling heat transfer in fusion divertor cooling channels.

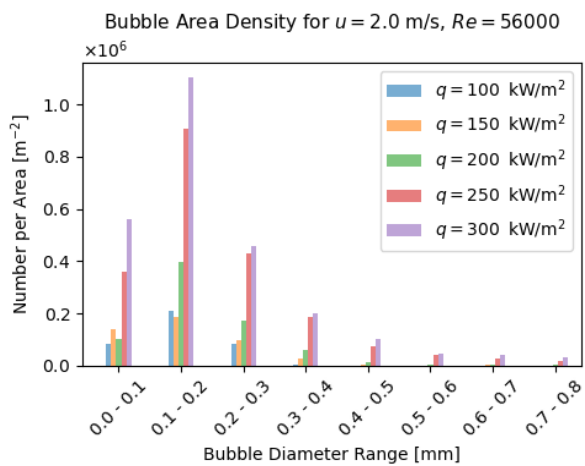

Fig. 8: Bubble size distribution for constant inlet velocity at three different heat fluxes.

## 5. Conclusions

In this work the motivation, key features, design process, scaling analysis, and some early results from the new FEDORA experiment are presented. Preliminary results demonstrate a proof-of-concept for studying boiling in a fusion divertor under laboratory conditions and represent a promising first step towards a comprehensive experimental campaign. Future work will focus on a detailed study of bubble dynamics and heat transfer mechanism under the scaled-down divertor cooling conditions. The experiment aims to provide a novel approach to addressing the gap in experimental measurements required for the development of computational fluid dynamics models suitable for divertor cooling conditions.

**Acknowledgements**

This research was conducted within the framework of the EUROfusion Consortium, funded by the European Union under the Euratom Research and Training Programme (Grant Agreement No. 101052200 — EUROfusion). The views and opinions expressed are solely those of the author(s) and do not necessarily represent those of the European Union or the European Commission. Neither the European Union nor the European Commission can be held responsible for them.

The financial support provided by the Slovenian Research Agency, grants P2-0026 and P2-0405 are gratefully acknowledged.